\documentclass[aps,tighten,floats]{revtex}
\usepackage[T1]{fontenc}
\usepackage{graphicx}
\IfFileExists{url.sty}{\usepackage{url}}{\let\url=\verb}

\makeatletter

\providecommand{\LyX}{L\kern-.1667em\lower.25em\hbox{Y}\kern-.125emX\@}
\newcommand{\noun}[1]{\textsc{#1}}

\makeatother

\begin{document}

\title{N\noun{uclear Aspects of the s- and n-Processes in Massive Stars}}

\author{T. Rauscher\thanks{
Thomas.Rauscher@unibas.ch
}}

\address{Dept. of Physics and Astronomy, University of Basel, Switzerland}

\author{A. Heger, S. E. Woosley}

\address{Dept. of Astronomy and Astrophysics, University of California, Santa Cruz,
CA, USA}

\author{R. D. Hoffman}

\address{Nuclear Theory and Modelling Group, Lawrence Livermore National Laboratory,
Livermore, CA, USA}

\maketitle
\begin{abstract}
In order to study the processes creating intermediate and heavy nuclei
in massive stars it is necessary to provide neutron capture cross sections
and reaction rates close to stability and for moderately unstable neutron-rich
nuclei. Furthermore, one has to know the efficiency of neutron-releasing
reactions in the main evolutionary phases of a massive star. We present
simulations of the nucleosynthesis in a 15 and 25 \( M_{\odot } \) star,
for the first time followed completely from main sequence hydrogen burning
until the type II supernova explosion including all nuclides up to Bi.
Theoretical reaction rates were calculated with the NON-SMOKER code, providing
a complete library of Hauser-Feshbach cross sections and rates for nuclear
and astrophysical applications. Experimental rates at stability were taken
from different sources. The impact of uncertainties in the rates on nucleosynthesis
are illustrated by two examples, the reactions \( ^{62} \)Ni(n,\( \gamma  \))\( ^{63} \)Ni
and \( ^{22} \)Ne(\( \alpha  \),n)\( ^{25} \)Mg.
\end{abstract}

\section{Introduction}

Stars above \( \simeq 10M_{\odot } \) are responsible for producing most
of the oxygen and heavier elements found in nature. Numerous studies have
been devoted to the evolution of such stars and their nucleosynthetic yields,
e.g., by Refs.~\cite{WW95} (in the following referred to as WW95) or \cite{TNH96}.
However, our knowledge of both the input data and the physical processes
affecting the evolution of these stars has improved dramatically in recent
years. Thus, it became worthwhile to improve on and considerably extend
the previous investigations of pre- and post-collapse evolution and nucleosynthesis.
We present the first calculation to determine, self-consistently, the complete
synthesis of all stable nuclides in any model for a massive star.

We employed a nuclear reaction network of unprecedented size in full stellar
evolution calculations. The network used by WW95, large in its days, was
limited to 200 nuclides and extended only up to germanium. Studies using
reaction networks of over 5000 nuclides have been carried out for single
zones or regions of stars with simplified networks, especially to obtain
the r-process, e.g., \cite{CCT85,fre99,kra93}, but ``kilo-nuclide'' studies
of nucleosynthesis in complete stellar models (typically of 1000 zones
each) have been hitherto lacking. With the continuous improvement in understanding
the input physics as well as in computer technology over the last several
years it became worthwhile and feasible to re-evaluate the nucleosynthesis
in massive stars using most recent physics and numerics.

In the following discussion we focus on the nuclear physics inputs and
their relevance to the production of intermediate and heavy nuclei. Due
to the increasing Coulomb barrier reactions involving neutrons play an
important role, either by directly producing the weak s-process component
and in a possible n-process (i.e.~an intermediate process between s- and
r-processes), or by providing the seed nuclei from which the \( \gamma  \)-process
produces proton-rich isotopes via photodisintegration processes.

\section{The Model}

\begin{figure}
\begin{center}
\includegraphics*[keepaspectratio=true,width=15cm]{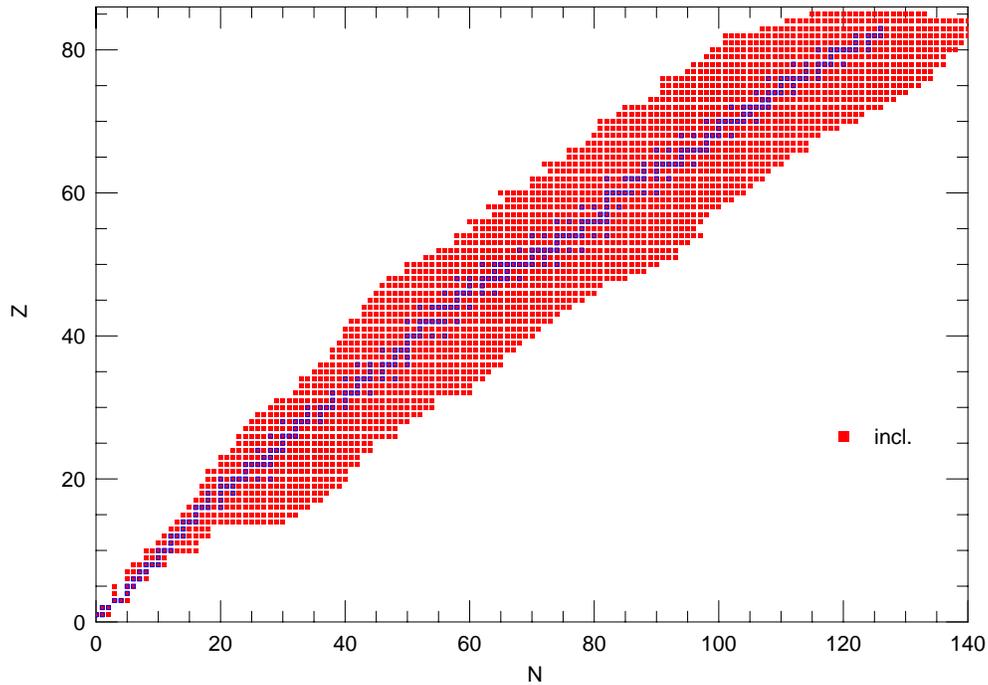}
\end{center}

\caption{\label{fig:net}Maximum size of reaction network during explosive 
burning. The line of stability is indicated.}
\end{figure}

Our calculations were performed using the stellar evolution code KEPLER
\cite{WW95,WZW78} with several modifications regarding astrophysical inputs
and the numerical treatment. The latter includes introduction of a novel
adaptive reaction network for nucleosynthesis considering all relevant
nuclides from hydrogen thru polonium at any given time; this amounts to
about 2350 nuclides during the explosion. The largest extension of the
network is shown in Fig.~\ref{fig:net}). For details see \cite{heg00,rau_apj}.
It is evident that there are many isotopes far off stability included for
which experimental information is scarce. This is even more true in more
exotic scenarios such as the r-process which involve isotopes close to
the dripline. (In our calculations we do not follow the proposed r-process
in the \( \nu  \)-wind emanating from the proto-neutron star shortly after
the collapse of the Fe core.)

As in previous studies \cite{WW95} the explosion was simulated using the
piston model. However, the piston procedure was slightly modified. We also
included the \( \nu  \)-process changing the light element abundances
during the explosive phase with the same cross sections as in WW95. We
updated the opacity tables (OPAL95) \cite{opal95} and considered
mass loss due to stellar winds \cite{mloss}. 
\begin{figure}
{\par\centering \resizebox*{1\columnwidth}{!}{\includegraphics{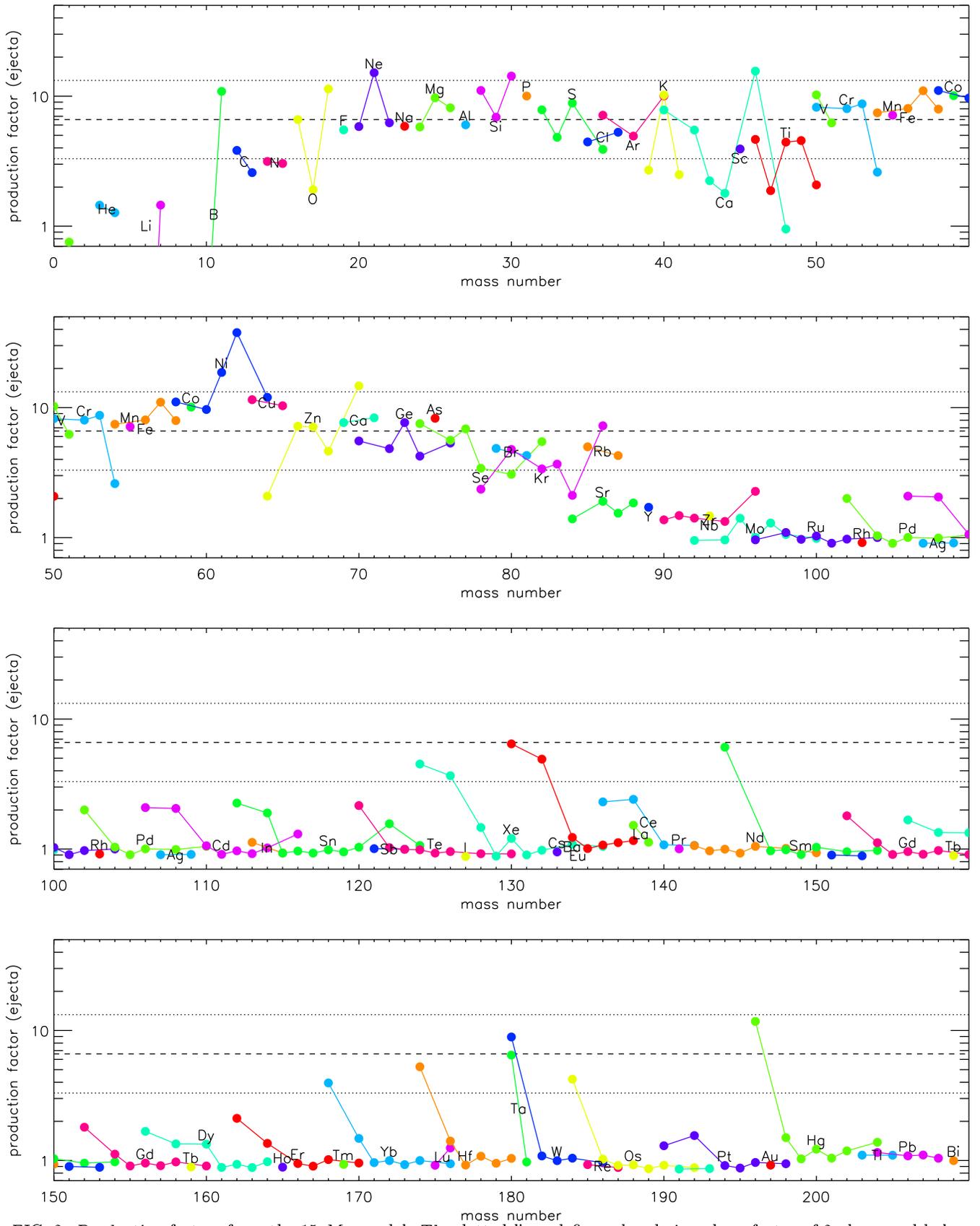}} \par}

\caption{\label{fig:abun}Production factors from the 15 \protect\( M_{\odot }\protect \) model. The dotted lines define a band given by a factor of 2
above and below the $^{16}$O production factor (dashed line).}
\end{figure}

\section{Theoretical and Experimental Nuclear Input}

Here, we only briefly summarize the most important nuclear inputs. For
details on the complete rate set, see Ref.\ \cite{rau_apj}.

\subsection{Statistical Model Rates}

As can be seen in Fig.\ \ref{fig:net}, a large part of our
network is comprised of unstable nuclei for which no experimental
information is available. Therefore we have to heavily rely on
theoretical predictions. It can be shown \cite{rtk97} that the majority
of those reactions proceeds via the compound nucleus reaction mechanism
and can be described in the framework of Hauser-Feshbach theory.
As the basis for the creation of our reaction rate set we used statistical
model calculations obtained with the NON-SMOKER code\cite{rt98,rt00}.
A library of theoretical reaction rates calculated with this code and fitted
to an analytical function -- ready to be incorporated into stellar model
codes -- was published recently \cite{rt00}. It includes about 30000
n-, p-, and \( \alpha  \)-induced
reaction rates for all possible targets from proton- to neutron-dripline
and between Ne and Bi, thus being the most extensive published library
of theoretical reaction rates to date. It also offers rate sets for a number
of mass models which are suited for different purposes. For the network
described here we utilized the rates based on the FRDM set. For other 
applications, nuclear cross sections for the whole mass range are published 
separately \cite{rt01} and are also available online 
\url|http://data.nucastro.org|.

Of special interest are $\alpha$ capture reactions on isospin self-conjugated
($N=Z$) target nuclei. The cross sections of such reactions are
suppressed by isospin effects and require special treatment in
theoretical models. This is described separately in Ref.\
\cite{rtgw00}.

\subsection{Weak Rates}

An important change of the weak interaction rates for $A\leq 40$ is
brought about by recent work \cite{lamar00}. For
the effects of those new rates on the stellar structure results, 
see Ref.\ \cite{hela00}.
The theoretical $\beta^-$ and $\beta^+$ rates of Ref.\
\cite{moe97} were also considered.
As a special case, we implemented a temperature-dependent $^{180\rm m}$Ta
decay \cite{end99}.
Updated neutrino loss rates were also used \cite{nuloss} but we did not
follow the $\nu$-process for nuclides with $Z$ or $N$ larger than 40.

\subsection{Experimental Rates}

The above described theoretical rates formed the backbone of the huge set
of reaction rates that was required to follow nucleosynthesis in our extended
reaction network. Whereever possible, the theoretical rates were substituted
by or supplemented with experimentally determined reaction and decay rates.

The experimental $\beta^-$, $\beta^+$, and $\alpha$ decay rates and
branching ratios given in Ref.\
\cite{NWC5} were implemented, including decays of the first excited
states.

Neutron capture rates were inferred from a compilation of 30 keV
Maxwellian Averaged Cross Sections \cite{bao2000} by renormalizing the
NON-SMOKER rates to the given values.

The most important reaction is the reaction 
$^{12}$C($\alpha$,$\gamma$)$^{16}$O. Its
value at 300 keV has still not be determined satisfactorily in
experiments (this is called {\em the} single most important
problem in experimental nuclear astrophysics) but a
variation of this rate impacts the C/O ratio obtained from He burning
and thus the stellar
structure and the total subsequent evolution and nucleosynthesis
of the star. The rate previously
used in WW95 was the one of Ref.\ \cite{CF88} multiplied by 1.7. Here,
the more recent evaluation of \cite{buc96} was used which yielded
$S(300)=146$ keV barn. The temperature dependence was chosen as
suggested by \cite{buc00}.

As a standard rate for \( ^{22} \)Ne(\( \alpha  \),n)\( ^{25} \)Mg we
used the rate of Ref.~\cite{kaepp} which is close to the lower limit of
the NACRE rate. This is also the rate used for the results shown in 
Fig.~\ref{fig:abun}. The uncertainties in this rate are discussed below
in Sec.\ \ref{sec:ne22}.

\section{Selected Important Nuclear Reactions}

The results from our 15 \( M_{\odot } \) model are shown in Fig.~\ref{fig:abun}.
The obtained abundances are not only sensitive to details in the stellar
physics (see e.g.~\cite{hoff99}) but also to nuclear physics. Nuclei beyond
the Fe peak can in general only be produced, directly or indirectly, via
neutron capture reactions which underlines the fact that it is important
to have a sound knowledge of the nuclear processes induced by neutrons
with energies of a few tens of keV. 

In this section two important reactions involving neutrons are discussed
in more detail. Both give rise to considerable uncertainties in the final
isotopic yields obtained in our model. They are not the only reactions
of special interest but due to the limited space we want to use these to
illustrate the difficulties one has to deal with when implementing nuclear
reaction rates in nucleosynthesis calculations. The first reaction -- \( ^{62} \)Ni(n,\( \gamma  \))\( ^{63} \)Ni
-- is important to model a specific detail in the abundance distribution
whereas the second reaction -- \( ^{22} \)Ne(\( \alpha  \),n)\( ^{25} \)Mg
-- is the main neutron source and therefore impacts the synthesis of all
intermediate and heavy nuclides.

\subsection{The Reaction \protect\( ^{62}\protect \)Ni(n,\protect\( \gamma \protect \))\protect\( ^{63}\protect \)Ni}

Examination of the production factors of the Ni isotopes in Fig.~\ref{fig:abun}
clearly shows a pronounced overproduction of \( ^{62} \)Ni in our calculation.
This especially catches the eye because all other nuclides are produced
in amounts nicely agreeing with the band defined via \( ^{16} \)O. A closer
look at the problem reveals a peculiar situation for the reaction \( ^{62} \)Ni(n,\( \gamma  \))\( ^{63} \)Ni
which destroys \( ^{62} \)Ni. For the results shown here, the reaction
rate was taken from Ref.~\cite{bao2000}. Interestingly, by comparison
with the previous version \cite{baoalt} of that reference it
is found that the rate has changed by about a factor of three between these
two compilations. The old value is higher by about a factor of three which
leads to increased destruction, sufficient to bring the \( ^{62} \)Ni
production down to a value comparable to the other nuclides in that
region in our calculation of the 15 \( M_{\odot } \) star.
\begin{figure}
\begin{center}
\includegraphics*[keepaspectratio=true,width=15cm]{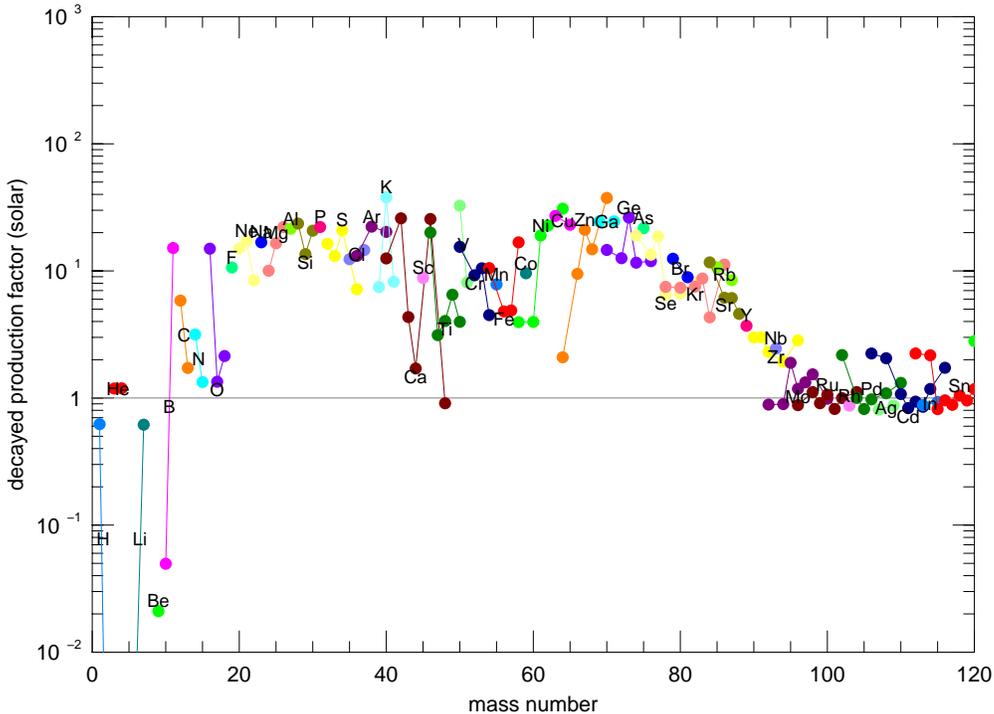}
\end{center}

\caption{\label{fig:nacre}Production factors in a 25 \protect\( M_{\odot }\protect \)
star with NACRE recommended rate for \protect\( ^{22}\protect \)Ne(\protect\( \alpha \protect \),n)\protect\( ^{25}\protect \)Mg.}
\end{figure}

Searching for the reason for the change in the \( ^{62} \)Ni(n,\( \gamma  \))\( ^{63} \)Ni
rate (cross section), it turns out that the new value is not based on a
more recent experiment but rather on a different extrapolation of the cross
section measured only at thermal neutron energies. The astrophysically
relevant neutron energies are in the range \( 20\leq E_{\mathrm{n}}\leq 40 \)
keV and the Maxwellian Averaged Cross Section at 30 keV is tabulated in
Refs.~\cite{bao2000,baoalt}. Since there is only a measurement of \( ^{62} \)Ni(n\( _{\mathrm{th}} \),\( \gamma  \))\( ^{63} \)Ni
using thermal neutrons n\( _{\mathrm{th}} \), an extrapolation to 30 keV
is needed. The older value uses a standard 1/\( v \) extrapolation, typical
for direct capture of an s-wave neutron. However, it was realized \cite{beerpriv}
that the 14.5 barn thermal neutron capture cross section of \( ^{62} \)Ni
is unusually high as compared to its neighboring isotopes, those showing
thermal cross sections which are lower by factors of \( 4-6 \). This lead
to the idea that the thermal \( ^{62} \)Ni capture cross section is enhanced due
to a subthreshold resonance in addition to the direct capture contribution
\cite{beerpriv}.
More specific, the excess value was attributed to the high energy tail
of a bound s-wave resonance at \( -0.077 \) keV. Since the resonance contribution
decays faster than 1/\( v \) to higher energies, at 30 keV only the contribution
of the direct term survives which is then lower by a factor of 3 than the
term assumed in the older rate extrapolation. Unfortunately, neither the
actual strength of the subthreshold resonance nor the neutron capture cross
section at 30 keV have been determined directly by an experiment. 

It would be audacious to infer from the discrepancies in our calculations
that the actual rate of \( ^{62} \)Ni(n,\( \gamma  \))\( ^{63} \)Ni
is lower by about a factor of 3 than the new value given in Ref.~\cite{bao2000}
and that the previously quoted value would be preferrable. However, in
the light of the experimental situation it is justified to direct the attention
of the experimentalists to that problem and to point out the importance
of neutron capture measurements in the astrophysically relevant energy
range instead of at thermal energies.

\subsection{The Reaction \protect\( ^{22}\protect \)Ne(\protect\( \alpha \protect \),n)\protect\( ^{25}\protect \)Mg}
\label{sec:ne22}

\begin{figure}
\begin{center}
\includegraphics*[keepaspectratio=true,width=15cm]{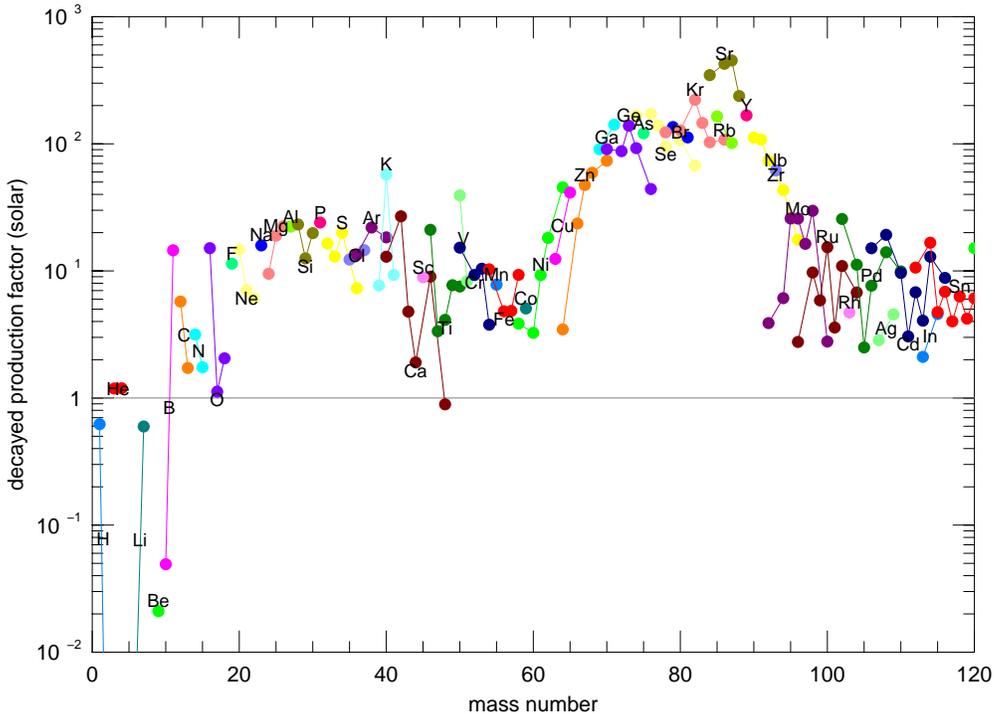}
\end{center}

\caption{\label{fig:nacrehi}Production factors in a 25 \protect\( M_{\odot }\protect \)
star with upper limit of the NACRE rate for \protect\( ^{22}\protect \)Ne(\protect\( \alpha \protect \),n)\protect\( ^{25}\protect \)Mg.}
\end{figure}

This reaction is the main neutron source in the type II supernova model.
It provides the neutrons to build up intermediate and heavy s-process nuclei
in the secondary s-process component originating from type II SN. These nuclei
are the seed nuclei for the \( \gamma  \)-process which produces proton-rich,
stable nuclei (p-nuclei) by photodisintegration (see e.g.~\cite{WW95,wh78,ray95}).
The produced p-nuclei can be nicely seen as the enhanced proton-rich
isotopes sticking out from the floor of weak s-process
abundances for $A>90$ in Fig.\ \ref{fig:abun}.
Thus, a change in this rate impacts a large number of nuclei, basically
all of the nuclides beyond Fe. 

It has also been speculated that a largely enhanced \( ^{22} \)Ne(\( \alpha  \),n)\( ^{25} \)Mg
rate could provide enough neutrons to make an r-process (e.g.\ Refs.\ 
\cite{tah79,cct80,bla81}). Despite of the
exponential dependence of the rate on stellar temperature, this would require
extreme temperature conditions, not found in our models. At most, we find
slightly increased temperatures during supernova shock front passage
through the He shell, leading
to a modest increase in the neutron flux over the standard s-process flux.
We call this an n-process which leads to a slight redistribution of abundances
but produces no r-process features and no significant total yields.

When trying to implement a proper rate for \( ^{22} \)Ne(\( \alpha  \),n)\( ^{25} \)Mg
one finds that the experimental errors are still considerably large. For
example, comparing the results when using the recommended value for that
reaction from the NACRE compilation \cite{NACRE} and the upper limit quoted
in the same compilation, we find that the production factors are higher
by factors of \( 10-30 \) for \emph{all} nuclides beyond Ga (Figs.~\ref{fig:nacre},
\ref{fig:nacrehi}). Recently, it was claimed that such a high value
could cure the underproduction of the p-process isotopes in the Mo-Ru
region which has been found in all stellar models \cite{costa00}.
Indeed, also those p-nuclei are produced more strongly, however, in a
consistent calculation like ours one clearly sees that such an
enhancement is ruled out by the fact that other nuclides are equally
enhanced. This not only limits the possible maximum value of the
$^{22}$Ne($\alpha$,n)$^{25}$Mg rate but also shows that another solution
has to be found for the production of the light p-nuclides in amounts
comparable to the more heavy p-nuclides.

Again, this shows the importance of measuring the cross sections in the
relevant energy range which is difficult in the case of \( ^{22} \)Ne(\( \alpha  \),n)\( ^{25} \)Mg.
Because of its importance in heavy element nucleosynthesis it is sometimes
called the second-most important reaction in nuclear astrophysics, after
the well-known reaction \( ^{12} \)C(\( \alpha  \),\( \gamma  \))\( ^{16} \)O.

\section{Summary}

The first calculation to self-consistently determine the complete
synthesis of all stable nuclides for massive stars was performed
with a reaction network of unprecedented size. Examples for the results
for a 15 $M_{\odot}$ and a 25 $M_{\odot}$ star were shown above. 

We find
a strong sensitivity of the production of intermediate and heavy nuclei
on the $^{22}$Ne($\alpha$,n)$^{25}$Mg rate which still remains to be
measured to satisfactory accuracy in the relevant energy range. 

As another example we pointed out that even neutron capture reactions on
stable isotopes, which are supposed to be well known, are still a source of 
large uncertainties.
A number of those reactions have only been measured at thermal neutron
energies and were then extrapolated to 30 keV. This is always
problematic because the energy dependence of the cross section can well
be different from the usually assumed 1/$v$ behavior, due to p-wave
capture and possible contributions of the tails of
resonances, both for resonances at higher excitation energies and
subthreshold resonances. It has to be strongly suggested that the
(n,$\gamma$) databases should be searched for such extrapolated cases
and that experimental efforts are undertaken to measure across a wider
range of energies. This would greatly improve the reliability of the
nuclear input and the credibility of the results of nucleosynthesis
calculations.

\end{document}